\documentclass[a4paper,11pt]{article}
\usepackage{pos}
\usepackage{graphicx}
\usepackage{epsfig}
\usepackage[leftcaption]{sidecap}
\usepackage{xcolor}
\usepackage{ulem}

\title{The Venusian Chronicles}

\author[a,b,c]{Luis A. Anchordoqui}
\author[d]{Carlos A. Garc\'{\i}a Canal}
\author*[d]{Sergio J. Sciutto}

\affiliation[a]{%
Department of Physics and Astronomy,  Lehman College, City University of
New York, NY 10468, USA.}
\affiliation[b]{%
Department of Physics, Graduate Center, City University
  of New York,  NY 10016, USA.}
\affiliation[c]{%
Department of Astrophysics, American Museum of Natural History, NY
 10024, USA.}
\affiliation[d]{%
Departamento de F\'{\i}sica and Instituto de F\'{\i}sica La Plata (CONICET), Universidad Nacional de La Plata (UNLP),
Calle 49 esq. 115, C. C. 67, 1900 La Plata, Argentina.}

\abstract{%
Venus' atmosphere --specifically its clouds buoyed up 40 to 60~km above the surface-- has long been suspected to encompass a biosphere where Earth-like living organisms could grow and flourish. This idea has been recently rekindled by the observation 
(signal-to-noise ratio of about 15$\sigma$) of a phosphine (PH$_3$) absorption-line profile against the thermal background from deeper, hotter layers of the atmosphere. There is a chance that this observation could be a sign of life, because the PH$_3$ gas observed on Earth originates mostly in decaying organic material. Furthermore, it has been shown that there is no other natural process on Venus that could otherwise produce the observed PH$_3$ absorption line. On the other hand, cosmic rays and the particle cascades they produce in the Earth's atmosphere are hazardous to living organisms, because the ionizing radiation produced in air showers can blow apart chemical molecules and disrupt biochemical processes, causing cells to die or undergo dangerous mutations. Compared to Earth, the hypothesized biosphere of Venus could be exposed to substantially more ionizing radiation. This is because Venus has no protective intrinsic magnetic field, orbits closer to the Sun, and the entire eventual habitable region lies in the clouds high in the atmosphere. Thereby, if the clouds  were sterilized there would be no reservoir of deeper life that can recolonize afterwards. In this communication we study the effects of  particle cascades in the venusian atmosphere using the AIRES simulation package properly configured. We show that the effects of cosmic radiation in the habitable zone would be comparable to those on the Earth's surface and so would not have any hazardous effect on possible venusian microorganisms. %
} 

\FullConference{7th International Symposium on Ultra High Energy Cosmic Rays (UHECR2024)\\
 17-21 November 2024\\
Malargüe, Mendoza, Argentina\\}

\begin{document}
\maketitle

%

\section{Chronology}

\label{intro}

{\it September 2020.} Millimeter-waveband spectra of Venus observed with the James Clerk Maxwell Telescope (JCMT) and the Atacama Large Millimeter/submillimeter Array (ALMA) came up with undeniable evidence (signal-to-noise ratio of about 15$\sigma$) of a phosphine (chemical formula PH$_3$) absorption-line profile against the thermal background from deeper, hotter layers of the atmosphere~\cite{Greaves:1}. Reanalyses of the JCMT and ALMA data to tackle three critiques questioning the bandpass calibration~\cite{Villanueva}, statistics on flase positives~\cite{Thompson}, and SO$_2$ contamination~\cite{Akins,Lincowski}  were immediately carried out and reported in~\cite{Greaves:2,Greaves:3,Greaves:4}. On top of that the PH$_3$ signal was also uncovered in archival data collected by the Pioneer Venus Large Probe Neutral Mass Spectrometer~\cite{Mogul}. The punch line of this discovery is that PH$_3$ is a biosignature gas associated with anaerobic ecosystems~\cite{Sousa-Silva}.

Potential pathways for phosphine production in a Venusian environment were investigated, with the conclusion that the observed PH$_3$ signal cannot be explained by conventional gas reactions, geochemical reactions, photochemistry, and other non-equilibrium processes~\cite{Bains}. The signal must then originate in some unknown geochemistry/photochemistry process, or else be the footprint of aerial microbial life. In this direction, a cycle for venusian aerial microbial life was developed
in~\cite{Seager}. The cycle builds upon dormant desiccated spores in the lower hot layers of the atmosphere. Updraft of spores transports them up to the habitable layer, where they ignite,   germinate, and become metabolically active microbes that grow and divide within liquid droplets. Eventually the droplets reach a size large enough to gravitationally
  settle down out of the atmosphere~\cite{Abreu}; higher temperatures and droplet
  evaporation trigger cell division and sporulation. The spores are
  small enough to withstand further downward sedimentation, remaining
  suspended in the lower haze layer
(a depot of hibernating microbial life) to restart the cycle.

{\it December 2021}. Motivated by these groundbreaking observations a series of focused astrobiology mission concepts to search for signs of life in Venus were proposed~\cite{Seager:2021m}. In particular, the Venus Life Finder mission will drop a small probe into the Venusian atmosphere with a single instrument (an autofluorescing nephelometer) to search for organic compounds.\footnote{{\tt https://venuscloudlife.com}} The probe will be carried to Venus aboard Rocket Lab's Photon spacecraft. The data collection phase will last roughly three to five minutes as the probe plummets from a height of 60 to $45~{\rm km}$, the potentially habitable region where the PH$_3$ signal was detected.

{\it June 2023}. Dedicated studies showed that single-chain saturated lipids with sulfate, alcohol, trimethylamine, and phosphonate head groups are resistant to sulfuric acid degradation at room temperature~\cite{Duzdevich,Seager:23}.

{\it November 2024}. Using the AIRES simulation package properly configured~\cite{Sciutto:2019jh}, we carried out a thorough analysis to ascertain the effects of cosmic particle cascades in the Venusian atmosphere. Compared to Earth, the hypothesized biosphere of Venus could be exposed to substantially more ionizing radiation. This is because Venus has no protective magnetic field, orbits closer to the Sun, and the entire eventual habitable region lies in the clouds high in the atmosphere. In contrast to the Earth's aerial microbial life, if the clouds   of Venus were sterilized there would be no reservoir of deeper life that can recolonize afterwards. In this communication we present a summary of our results. A full discussion will be provided elsewhere~\cite{Anchordoqui:25}.

\begin{SCfigure}
\includegraphics[width=0.5\linewidth]{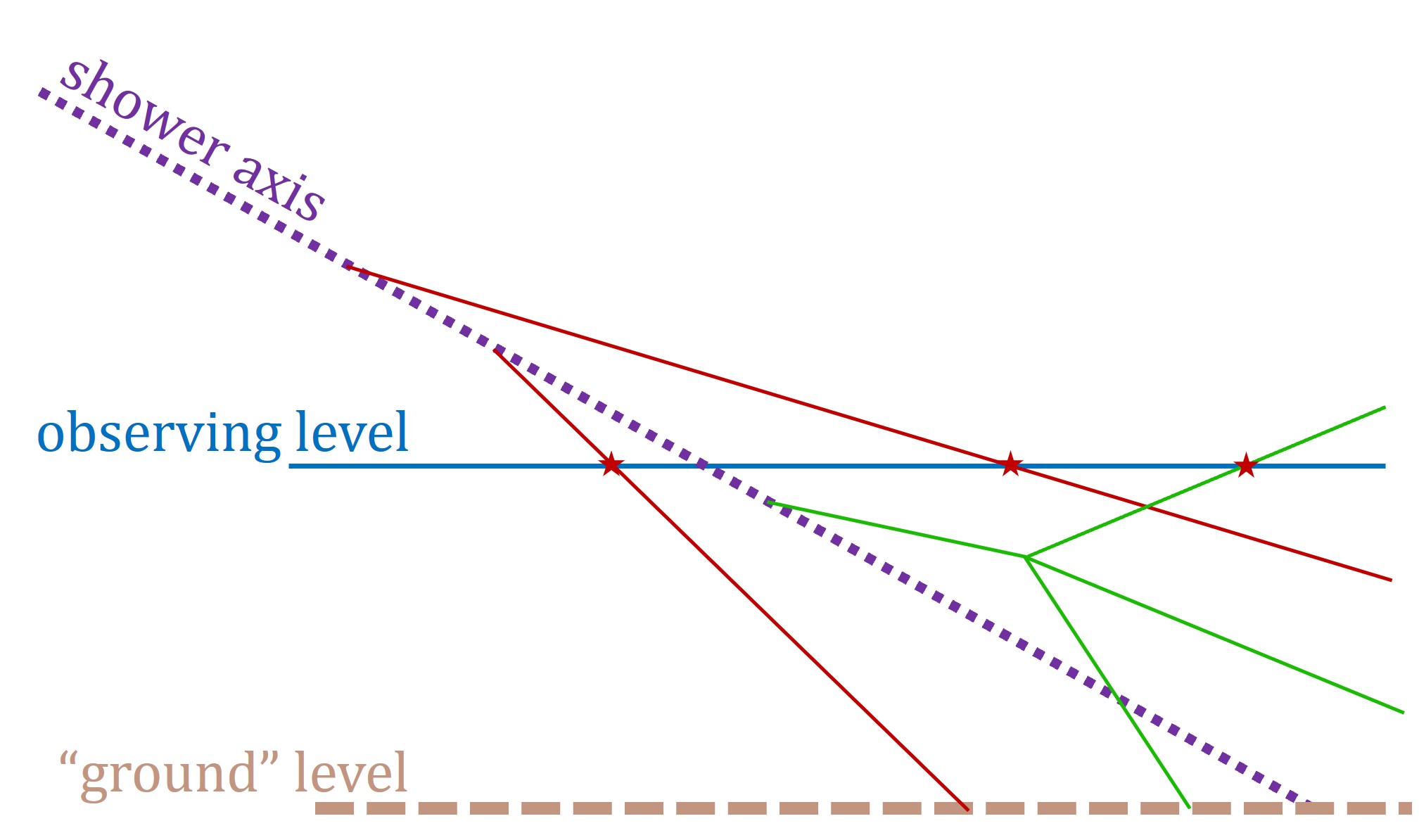}
\caption{Schematic representation of a shower tresspassing a given observing level. Noctice that the ground level used in the simulations must be located below the observing level to allow for proper propagation of upgoing particles.}
\label{fig:1} 
\end{SCfigure}

\section{A Compendious of Cosmic Ray Showers in Venus}

A simulation study of the energy released in the form of hazardous
radiation by cosmic ray showers developing within the Venusian
atmosphere was carried out using the AIRES simulation code~\cite{Sciutto:2019jh}. The
high-energy hadronic interactions were handled by SIBYLL 2.3d~\cite{Riehn:2019jet}, whereas
the low-energy interactions were simulated with the ``built-in'' model
that comes with AIRES. We note in passing that there are no major differences between the predictions of the hadronic interaction models implemented in AIRES at the energies relevant to this study.  This is not the first time that AIRES was used
to simulate particle fluxes in the atmosphere. Indeed, e.g.,
the simulated atmospheric muon flux with AIRES
successfully reproduced CAPRICE measurements~\cite{CapriceFlux}. This is an important
validation of AIRES and of its low-energy hadronic model to
accommodate the number of muons. For the problem at hand, the
simulation is more complex than in the study of CAPRICE data. This is
because the simulation must account for the flux of all secondary
particles (not only muons), and in all directions (i.e., the total
flux at a given altitude). The simulation of the total particle flux
in all directions is possible because AIRES propagates all particles,
regardless of the direction in which they move, see Fig.~\ref{fig:1}.
The mass of simulations run to generate the particle flux data at
different levels exceeds 1 billion showers. As far as we are aware, no
previous study of cosmic ray showers has such a colossal statistics.

\begin{figure}[htb!]
  \begin{minipage}{0.495\textwidth}
      \includegraphics[width=0.99\linewidth]{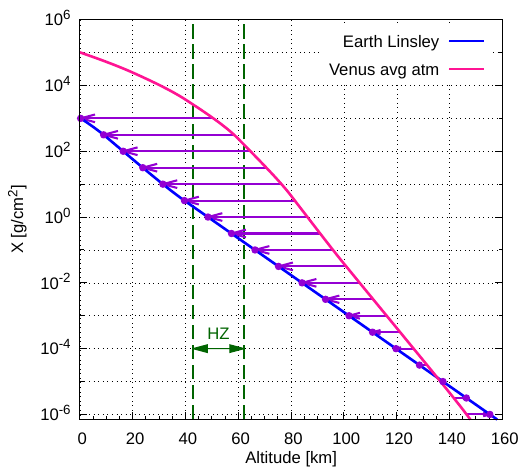}
   \end{minipage}
  \begin{minipage}{0.495\textwidth}
    \includegraphics[width=0.99\linewidth]{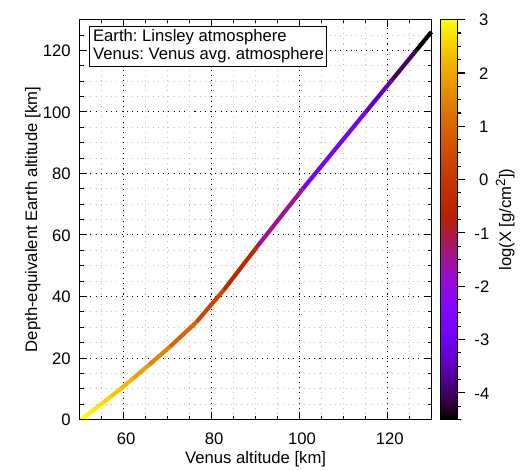}
  \end{minipage}         
  \caption{Comparison of Venus and Earth atmospheric profiles (left) and depth-equivalent Earth altitude versus Venus altitude (right). Data of Venus atmosphere taken from~\cite{VenusAtmosphereSeiff}; Earth atmosphere represented with Linsley's model~\cite{Sciutto:2019jh,Hillas:1997tf}.  The vertical green dashed lines at the left panel indicate the extremes of the ``Habitable Zone'' (HZ) of reference \cite{VenusBiosphere1} (venusian altitudes in the range 42.8 km to 62.2 km).}
  \label{fig:2}
\end{figure}

We begin the discussion with a characterization of the Venusian atmosphere.
In Fig.~\ref{fig:2} we show a comparison of the atmospheric profiles
of Venus and Earth. In Fig.~\ref{fig:2}a we show the atmospheric depth as a
function of altitude, measured in
km from the surface of the corresponding planet. The atmospheric
profile of the Venusian atmosphere is a multilayer model implemented
in AIRES to fit the experimental data~\cite{VenusAtmosphereSeiff}. In
Fig.~\ref{fig:2}b we introduce the concept of equivalent Earth altitude,
by showing the altitude on Earth of a point that has the same
atmospheric depth as a given Venusian altitude. The equivalent Earth
altitude can only be defined for those points in the Venusian
atmosphere whose atmospheric depth is less than the Earth's
atmospheric depth at sea level (approximately $1,000~{\rm g/cm}^2$). The altitude dependence of the atmospheric pressure is described by the US standard atmosphere as parametrized by Linsley~\cite{Sciutto:2019jh,Hillas:1997tf}. The equivalent Earth altitude can be calculated from 50.3~km of Venusian altitude. The configuration of AIRES for simulations with Venus atmosphere includes also adjusting its composition (96.5\% carbon dioxide, 3.5\% nitrogen), and hence the radiation length. An adjustment of the planet's radius is also necessary, together with disabling the intrinsic magnetic field (in the case of simulations at the Earth, a geomagnetic field intensity of $\sim 32\, \mu{\rm T}$ is used).

The clouds of Venus surround the entire planet, yielding a high
planetary albedo $\sim 0.8$~\cite{Marov}. The starting altitude of this thick cloud cover is $\sim 47$~km above the
surface (the temperature at this position is around $100^\circ$C) and extends up to over 70~km. The distribution aerosol particles drifting inside the
clouds can be classified according to their size into three substrata:  lower (47.5 to 50.5~km), middle (50.5 to 56.5~km), and upper (56.5 to 70~km
altitude); the
smallest type-1 droplets (with a radius of $0.2~\mu{\rm m}$) and type-2 droplets
(with a radius of 1 to 2~$\mu{\rm m}$) nest in all three substrata, whereas the largest type-3 droplets (with a radius of
$4~\mu{\rm m}$) only live in the middle and lower cloud layers~\cite{Knollenberg}. We will refer to the 47 to 58~km region as the core of the habitable zone, and use a more generous range to describe the habitable zone~\cite{VenusBiosphere1}. In the $47  \lesssim  {\rm altitude/km} \lesssim 58$ range, the temperature is $0 \lesssim T/^\circ{\rm C} \lesssim 100$, the pressure is $\sim 1$~atm, and thereby the existence of life is most likely~\cite{Seager,VenusBiosphere2}.
As can be seen in Fig.~\ref{fig:2}, the upper boundary of the habitable zone
 is at $\sim 62$~km altitude, and hence it receives less than $200~{\rm g/cm^2}$
shielding depth against cosmic ray showers. This is far less than the
biosphere on the surface of the Earth, which is beneath $1,033~{\rm
  g/cm^2}$. Nevertheless, the atmospheric grammage in the middle substratum is
$\sim 1,000~{\rm g/cm^2}$, insinuating that the effects of cosmic ray cascades  would be similar to those on the
Earth's surface. It is this that we now turn to study.
\begin{figure}[htb!]
  \begin{minipage}{0.495\textwidth}
    \begin{centering}
      \includegraphics[width=0.99\linewidth]{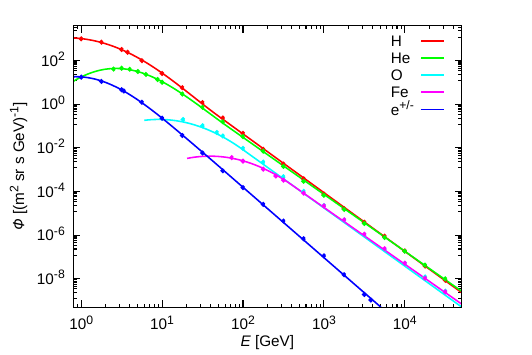} \\
      \includegraphics[width=0.99\linewidth]{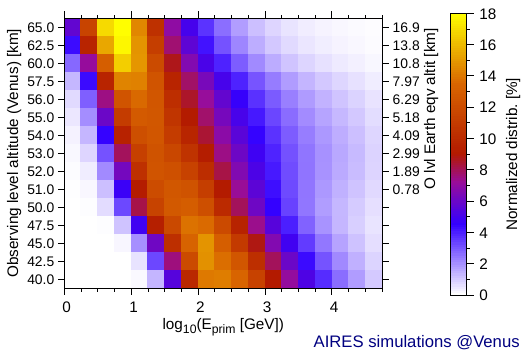}
    \end{centering}
   \end{minipage}
  \begin{minipage}{0.495\textwidth}
    \includegraphics[width=0.99\linewidth]{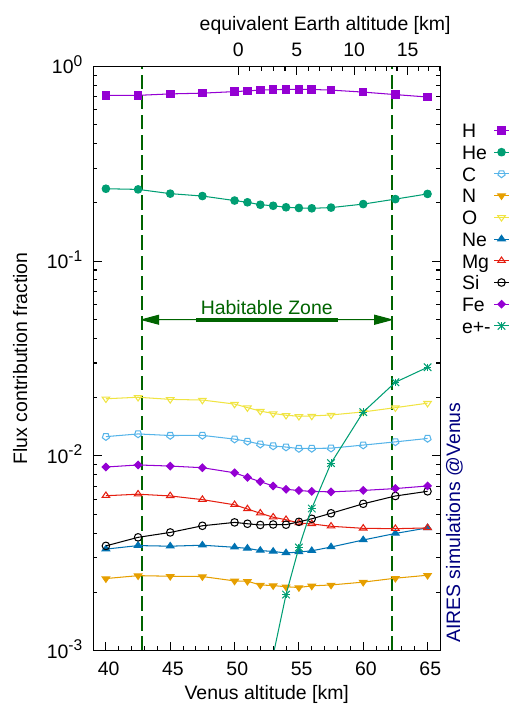}
  \end{minipage}         
  \caption{{\bf(a)} {\it (upper left)\/} TOA primary fluxes used in AIRES air shower simulations. The dots correspond to data from the most recent review from reference~\cite{PDG2024}, while the solid lines represent the corresponding fits to the data that were actually used in the simulations. For clarity, only the most representative primaries are displayed (our simulations include: H, He, C, N, O, Ne, Mg, Si, Fe, $e^+e^-$). {\bf(b)} {\it (lower left)\/} Distribution of the contribution of primaries of different energies to the total flux at all the altitudes considered in the present study (Normalized to 100\% at each observing level). {\bf(c)} {\it (right)\/} Contributions of different primaries to the total flux as a function of altitude. The vertical green dashed lines at the left panel indicate the extremes of the ``habitable zone'' of reference~\cite{VenusBiosphere1}, and the thick segment of the horizontal double arrow marks the core habitable zone (see text).}
  \label{fig:3}
\end{figure} 

\begin{figure}[htb!]
  \begin{minipage}{0.495\textwidth}
    \includegraphics[width=0.99\linewidth]{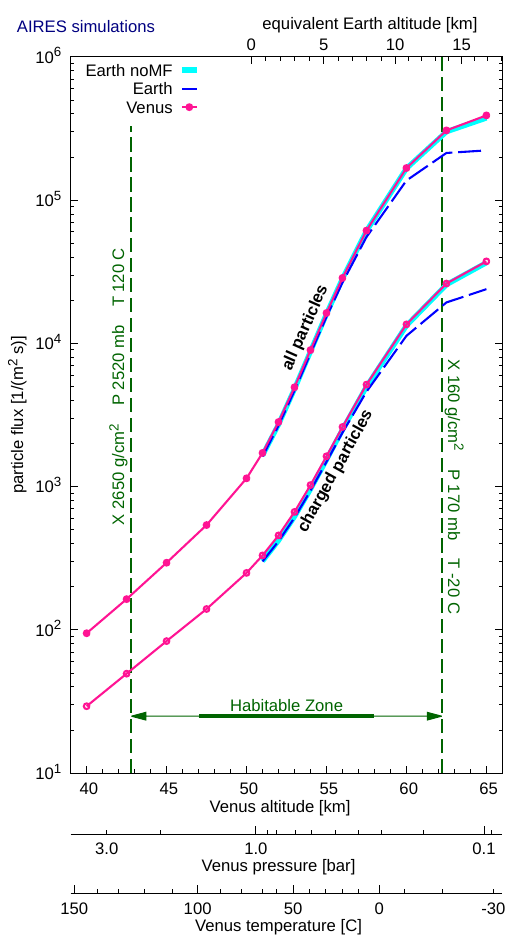}
  \end{minipage}
  \begin{minipage}{0.495\textwidth}
    \begin{centering}
      \includegraphics[width=0.99\linewidth]{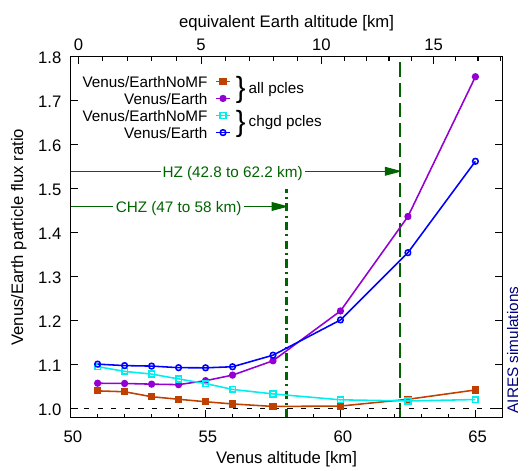} \\
      \includegraphics[width=0.99\linewidth]{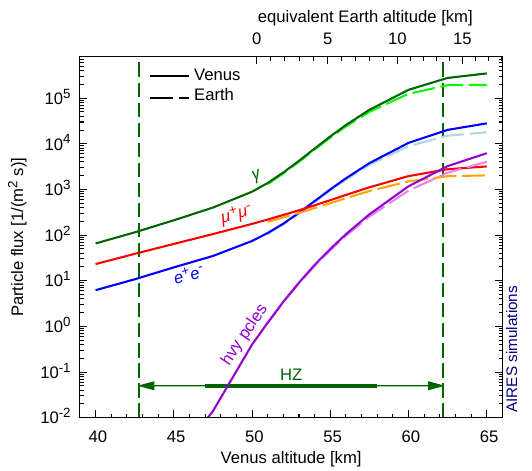}
    \end{centering}
   \end{minipage}
  \caption{{\bf(a)} {\it (left)\/} All and charged particle fluxes versus altitude, showing the cases of Venus atmosphere, and Earth atmosphere with and without taking into account the geomagnetic deflections of particles. Horizontal scales for atmospheric pressure and temperature are also included. The vertical green dashed lines indicate the extremes of the ``Habitable Zone'' of reference \cite{VenusBiosphere1}, together with the venusian average atmosphere parameters at the corresponding zone ends (atmospheric depth, pressure, and temperature). The thick segment of the horizontal double arrow marks the core habitable zone (see text). {\bf(b)} {\it (upper right)\/} Venus/Earth total flux ratios. {\bf(c)} {\it (lower right)\/} Various secondary particles fluxes versus altitude.}
  \label{fig:4}
\end{figure} 

An essential hypothesis of our study is the intensity of the cosmic ray flux at the top of Venus' atmosphere. We assume that this intensity is equal to the intensity of the cosmic ray flux at the top of Earth's atmosphere. The relevant parameters for the shower simulations are: {\it (i)}~Environmental parameters (geometry, atmosphere, magnetic field, already discussed). {\it (ii)}~The primary composition with abundances, which is  illustrated by representative species in Fig.~\ref{fig:3}a. {\it (iii)}~The primary energy, taken between 1~GeV and 50~TeV (enough to cover the region of relevance for
the flux calculation; see Fig.~\ref{fig:3}b). {\it (iv)}~The observation levels from 40 to 65~km Venusian altitude, and from 0 to ~17 km Earth altitude. Now, how credible is our essential hypothesis? Although it seems obvious, we first note that the fluxes of Galactic cosmic rays reaching the Earth and Venus are the same, whereas the fluxes of solar cosmic rays are different. The baryonic cosmic rays that have influence on the solar cycle are those with primary energy $\lesssim$ GeV. As we can see in Fig.~\ref{fig:3}b, the contribution of cosmic rays near 1~GeV to the flux of secondary particles is negligible (reaching 4\% above 60~km). Indeed, at all altitude levels the relevant primary flux that determines the local secondary particle flux corresponds to energy bins well away from the range influenced by the solar cycle. One may now wonder, what is the contribution of baryonic cosmic rays to the total flux? To address this question, in Fig.~\ref{fig:3}c we show the contribution of different primaries to the total secondary particle flux. It is straightforward to verify that primary protons and helium nuclei are responsible for around 96\% of the total flux of secondary particles, at any of the altitudes considered.
Most of the remaining primary nuclei contribute less than 2\%. Now, in our hypothesis the contribution of electron and positrons producing particle cascades in Venus is $\sim 10^{-3}\%$ at 50~km, and it reaches $\sim 3\%$ at 60~km. A point worth noting at this juncture is that the primary flux of electrons is likely to be significantly different from that on Earth, because its dominant contribution comes directly from the sun, and is largely dependent on the magnetic field, both solar and planetary. On the assumption that the flux falls like the inverse square law, we find the flux at Venus would be increase by only a factor of two. However, the electron flux at Venus could 
actually be larger due to intricate behavior of the interplanetary magnetic field. In summary, the hypothesis that the cosmic ray fluxes in Venus and Earth are similar is reasonable for baryons, but not so clear for electrons. However,  electrons do not have a great influence unless the flux is increased in Venus by several orders of magnitude. Therefore, we can conclude that the primary flux of electrons has no influence on biological effects.

Our main results are encapsulated in Fig.~\ref{fig:4}. One can check by inspection of Fig.~\ref{fig:4}a that the secondary particle fluxes in the core of the ``habitable zone'' of Venus are not
significantly different from those we can measure on Earth (at
equivalent altitudes).  For a direct comparison, in Fig.~\ref{fig:4}b we show the ratio between the flux on Venus over
the flux on Earth (with and without geomagnetic field). Note that if we compare the flux in Venus to that on Earth assuming zero terrestrial geomagnetic field, then the Venusian and terrestrial fluxes are practically the same. In the upper edge of the habitable zone
(altitudes above
about 60~km) the fluxes on Venus are slightly greater ($\sim 40\%$) due to the lack of a Venusian intrinsic magnetic shield. In Fig.~\ref{fig:4}c we compare the quantity and the type of composition of the secondaries present in the habitable zone of Venus and the Earth. While we cannot conclude what the effect on the spores will be, we can argue that the way the radiation would affect spores populating the clouds of Venus is quite similar to how it would affect microbes on Earth at a similar equivalent altitude. 

We end with an observation. We take notice that the flux of protons can be enhanced during exceptionally strong solar flares. As noted in~\cite{Herbst}, such an increase of the flux would lead to an increase of the ionization in the core of the habitable zone of Venus. However, we note that a roughly equivalent proton flux will also increase the ionization at the Earth's surface. At Earth-Venus comparable depth-equivalent altitudes where microorganisms can grow we expect moderate differences in the flux of charged particles.
An investigation along these lines is obviously important to be done~\cite{Anchordoqui:25}.

\section{Take Home Message}

We have studied the major characteristics of cosmic ray showers developing in the Venusian atmosphere,  taking profit of the ductility of the  AIRES simulation package to cover a large range of energies. Despite the fact that the particles in the cascade have different propagation properties than in cosmic ray air showers (e.g., there is no intrinsic magnetic field in Venus) we have shown that there are no relevant differences between the number of particles and the number of charged particles (which drive the ionization in the atmosphere) at comparable depth-equivalent altitudes where life can flourish. We can then conclude that that effects of cosmic radiation in the habitable zone of Venus would be comparable to those on the Earth’s surface, and so would not have any hazardous effect on possible Venusian microorganisms.\\

\section*{Acknowledgments}
We are indebted to Beatriz Garc{\'\i}a for useful comments about the cosmic ray flux in the solar system. S.J.S and C.G.C. are partially supported by ANPCyT. The work of L.A.A. is supported by the U.S. National Science Foundation (NSF Grant PHY-2412679); he thanks the Theoretical Physics Laboratory at the UNLP in Argentina for hospitality during completion of this work.



\end{document}